# A non-invasive dry-transfer method for fabricating mesoscopic devices on sensitive materials

*Zhongmou Jia,* [1,2] *Yiwen Ma,* [1,2] *Zhongchen Xu,* [1,2] *Xue Yang,* [1,2] *Jianfei Xiao,* [1,2] *Jiezhong He,* [1,2] *Yunteng Shi,* [1,2] *Zhiyuan Zhang,* [1,2] *Duolin Wang,* [1,2] *Sicheng Zhou,* [1,2] *Bingbing Tong,* [1,3] *Peiling Li,* [1,3] *Ziwei Dou,* [1] *Xiaohui Song,* [1,3] *Guangtong Liu,* [1,3] *Jie Shen,* [1] *Zhaozheng Lyu,* [1,3] *Youguo Shi,* [1,2,4,a)] *Jiangping Hu,* [1,2,5,a)] *Li Lu,* [1,2,3,a)] *and Fanming Qu* [1,2,3,a)]

**AFFILIATIONS**

[1] Beijing National Laboratory for Condensed Matter Physics, Institute of Physics, Chinese Academy of Sciences, Beijing 100190, China

[2] University of Chinese Academy of Sciences, Beijing 100049, China

[3] Hefei National Laboratory, Hefei 230088, China

[4] Songshan Lake Materials Laboratory, Dongguan, Guangdong 523808, China

[5] New Cornerstone Science Laboratory, Shenzhen 518054, China

a) Authors to whom correspondence should be addressed: ygshi@iphy.ac.cn; jphu@iphy.ac.cn; lilu@iphy.ac.cn; fanmingqu@iphy.ac.cn

ABSTRACT. Many materials with novel or exotic properties are highly sensitive to environmental factors such as air, solvents, and heat, which complicates device fabrication and limits their potential applications. Here, we present a universal submicron fabrication method for mesoscopic devices using a dry-transfer technique, tailored specifically for sensitive materials. This approach utilizes PMMA masks, combined with a water-dissoluble coating as a sacrificial layer, to ensure that sensitive materials are processed without exposure to harmful environmental conditions. The entire fabrication process is carried out in a glove box, employing dry techniques that avoid air, solvents, and heat exposure, culminating in an encapsulation step. We demonstrate the utility of this method by fabricating and characterizing $K_2Cr_3As_3$ and $WTe_2$ devices, a one- and two-dimensional material, respectively. The results show that our technique preserves the integrity of the materials, provides excellent contact interfaces, and is broadly applicable to a range of sensitive materials.

TEXT. Micro-fabrication techniques are fundamental to the measurement and application of materials, providing a means to study materials at the mesoscale. The primary goal of micro-fabrication is to construct devices with high performance and minimal impact on material properties. Over recent decades, the growth of material libraries has expanded dramatically, with the continuous synthesis of new materials and the development of novel device structures [1-7]. A vast number of materials now have the potential to be exfoliated into two-dimensional (2D) sheets, making them ideal candidates for next-generation devices [8, 9]. This has also facilitated the cleavage, exfoliation, and transfer techniques of samples, such as chemical-etchant assisted wet transfer, water-soluble layer-based transfer, thermal-release transfer, electrochemical bubble transfer and metal-assisted transfer [1-7,10,11]. Similarly, device structure fabrication technologies have also developed rapidly, such as conventional lithography and thin-film deposition techniques, nanoimprint lithography, and transfer electrode technology [12,13]. However, there are certain limitations in using these technologies to fabricate devices on sensitive materials. Conventional lithography requires solvents and heating, causing damage to sensitive materials. Nanoimprint lithography generates mold-release agents and thermo-mechanical stress during embossing, resulting in interfacial contamination and fracture. Transferred-electrode methods for vdW contacts are constrained by thermal-release temperature promoting metal diffusion and interfacial stress. Therefore, the electrical transport measurements of sensitive materials are greatly limited to a large extent. Currently, measurements of sensitive materials fall into two broad categories: non-contact and contact-based techniques. Non-contact methods, such as nuclear magnetic resonance, Raman scattering, powder X-ray diffraction, and tunnel diode resonators [14-20], are invaluable for investigating material properties without physical interaction. On the other hand, contact methods involve directly interfacing with the material, for example, encapsulating graphene between boron nitride to fabricate high-mobility devices [1] or attaching wires onto materials with silver [17]. These contact methods are often carried out within glove boxes to prevent degradation due to exposure to air or solvents [21, 22]. While both approaches have seen significant success on investigating sensitive materials, they are limited in several ways especially for those extremely sensitive materials, including size constraints, difficulty in probing submicron devices, and challenges in achieving direct, high-quality electrical contacts. Given these limitations, there remains a pressing need for a mesoscopic-device fabrication method that avoids exposure to air, solvents, and heating, allowing sensitive materials to be processed in a controlled, non-invasive manner.

In this work, we propose an all-dry approach that leverages PMMA masks to fabricate high-quality mesoscopic devices involving both electrodes and encapsulation layers, without the need for baking or chemical exposure. The free-standing PMMA mask, with pre-defined electrode patterns fabricated using a standard electron beam lithography technique, is obtained by dissolving a sacrificial layer. This mask is then transferred onto a PDMS stamp, and subsequently aligned with and attached to the target material under an optical microscope. A metal film is then deposited, followed by peeling off the mask, leaving behind the electrode pattern. The materials are finally encapsulated in a similar manner.

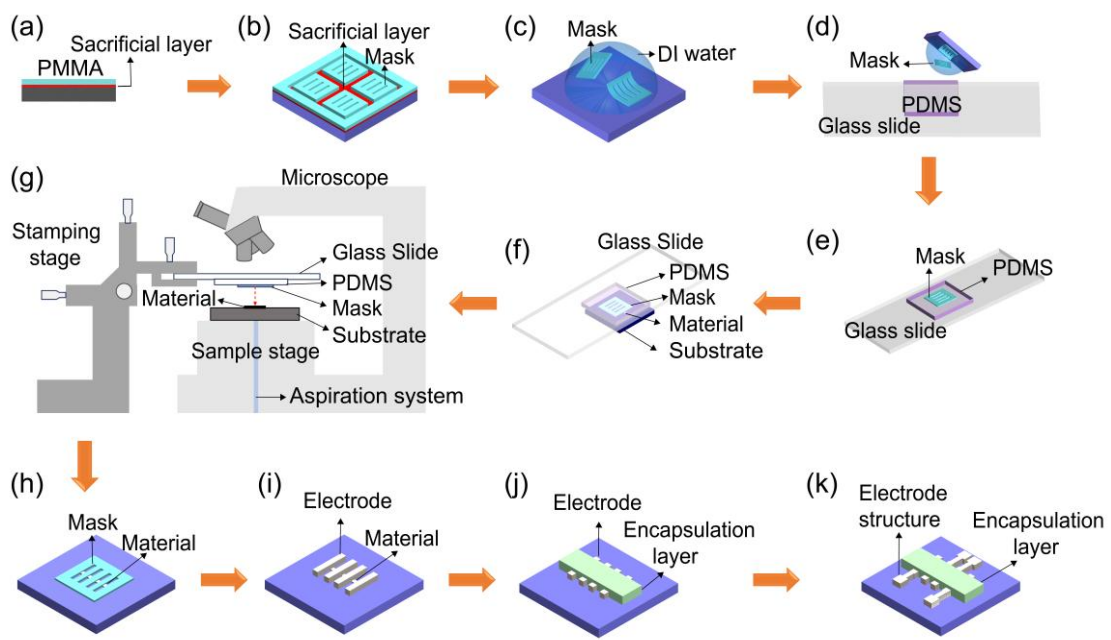

**FIG. 1.** Schematic diagram of the non-invasive dry-transfer method for sensitive material device fabrication. (a) Spin-coating of a sacrificial layer and PMMA layer on Si/SiO$_2$ substrate. (b) Electron-beam lithography patterning to form a PMMA mask array. (c) Dispensing deionized (DI) water droplet onto the substrate. DI water permeates mask-substrate gaps, dissolving the sacrificial layer and causing PMMA masks to float on the droplet surface. (d) Inverting the substrate to gently contact pre-prepared PDMS/glass slide, transferring the droplet with masks. (e) Removing the droplet via N$_2$ blowing, leaving masks on PDMS. (f), (g) Aligning and angular adjustment of the mask relative to the target material. (h) Releasing the mask onto the material, followed by metal deposition. (i) Peeling off the metal film and mask using tape to form electrodes. (j) Repeating steps (a)-(i) to fabricate an encapsulation layer (e.g., Al$_2$O$_3$). (k) Extending electrodes via standard micro-fabrication for wire bonding and measurement.

The fabrication of the PMMA mask begins with ultrasonic cleaning of Si/SiO$_2$ (300 nm) substrates in acetone and isopropanol. A 40-nm-thick water-soluble conductive coating (5090.02) is then spin-coated as the sacrificial layer, followed by PMMA resist deposition [Fig. 1(a)]. For larger masks, the sacrificial layer and PMMA layer should be thicker by reducing the spin speed or using higher-concentration PMMA. After electron-beam lithography patterning [Fig. 1(b)], DI water is dropped onto the substrate. Water dissolves the sacrificial layer through mask-substrate gaps, causing the mask to float on the surface of the water droplet [Fig. 1(c)]. The substrate is inverted and gently touched to the PDMS, transferring the water droplet with the mask onto the PDMS [Fig.

1(d)]. The PDMS stamp is a thin layer of commercially available viscoelastic material, which is adhered to a glass slide for easy handling. Subsequently, a nitrogen gun is used to blow away the water droplets, leaving the masks on PDMS surface [Fig. 1(e)]. To ensure the masks is free of moisture, we placed them in a vacuum environment for several minutes, allowing any remaining moisture to evaporate. Then, the assembly was transferred into the glovebox. In the glovebox, the mask is aligned with and attached to the material, and then the assembly is transferred to a vacuum deposition system connected to the glovebox for metal deposition. The precise steps of mask releasing shown in Fig. 1(f), (g), (h) can be summarized into the following six points: (1) Adsorbing the material substrate onto the sample stage. (2) Inverting the PDMS/glass slide onto the stamping stage, resulting in the following vertical order from top to bottom: glass slide-PDMS-mask-material. (3) By adjusting the positions of the sample stage and stamping stage, the mask is positioned above the material. (4) Rotating the sample stage for device orientation selection. (5) Lowering the stamping stage until the mask contacts the material. (6) Raising the stamping stage to lift the PDMS, leaving the mask adhered to the material. After metal deposition, the metal film and mask are peeled off using tape, forming electrodes [Fig. 1(i)]. Encapsulation layers (e.g., $Al_2O_3$) are fabricated using the same procedure [Fig. 1(a)-(i)]. Fig. 1(j) shows the final encapsulated device, where the PMMA mask has been removed, leaving the encapsulated material with uncovered electrode segments. The electrodes can be extended using standard micro-fabrication techniques, enabling wire bonding and further device integration for measurement [Fig. 1(k)]. Noteworthy, to eliminate heating, the photoresist is cured in vacuum instead of baking in electrode extension steps.

The key advantages of our method are summarized as follows. 1) No contamination: The method prevents sensitive materials from contacting photoresist, solvents, developers, fixers, and air, effectively eliminating organic residues and ensuring higher-quality contacts with the material. 2) No heating: The electrode pattern is predefined using a PMMA mask, which eliminates the typical requirement of photoresist baking on the materials. 3) No thickness limitation: Unlike traditional methods, our technique does not impose restrictions on the material thickness, making it suitable for a wide range of materials. 4) Sub-micron accuracy: The method achieves fabrication accuracy at the sub-micron level, offering significant advantages in device design, especially for applications that require highly precise and miniaturized structures.

To ensure successful fabrication, it is worth noting that: (1) The thickness of the PMMA mask should be selected according to the size: the larger the mask, the thicker the PMMA film layer. (2) When preparing the mask for electrodes, the subsequent extension range of the electrodes must be considered. A 600-μm mask is recommended. (3) After metal deposition, the substrate must be firmly fixed to ensure a single successful peeling during the mask peel-off process using tape. Otherwise, residue may remain on the material surface. (4) The position of the material on the substrate is critical: the closer to the edge of the substrate, the higher the precision and success rate.

To demonstrate the capabilities of our method, we fabricated devices from two distinct materials: $K_2Cr_3As_3$, a one-dimensional (1D) material, and $WTe_2$, a 2D layered material, ideal candidates for testing the efficacy of our non-invasive fabrication technique. $K_2Cr_3As_3$, which is extremely sensitive and has previously seen limited microscopic electrical measurements [23], was fabricated into devices using our dry-transfer method. We measured the superconducting transition temperature ($T_c$~5 K) and observed a linear temperature dependence of resistivity over a wide range (5-300 K), which is indicative of non-Fermi liquid behavior and consistent with previous reports on bulk materials [24]. Additionally, we found that our $K_2Cr_3As_3$ devices retained their superconductivity even under a vertical magnetic field of 12 T, aligning with findings from prior studies on bulk materials [25, 26]. For the 2D material $WTe_2$, we designed and fabricated a mesoscopic Hall-bar device with a contact resistance of only several ohms. The device exhibited negative longitudinal magnetoresistance when the magnetic field was applied parallel to the current direction, which is a signature of its topological property. Additionally, we investigated the field and angular dependence of the magnetoresistance, which further confirmed the unique electrical properties of $WTe_2$.

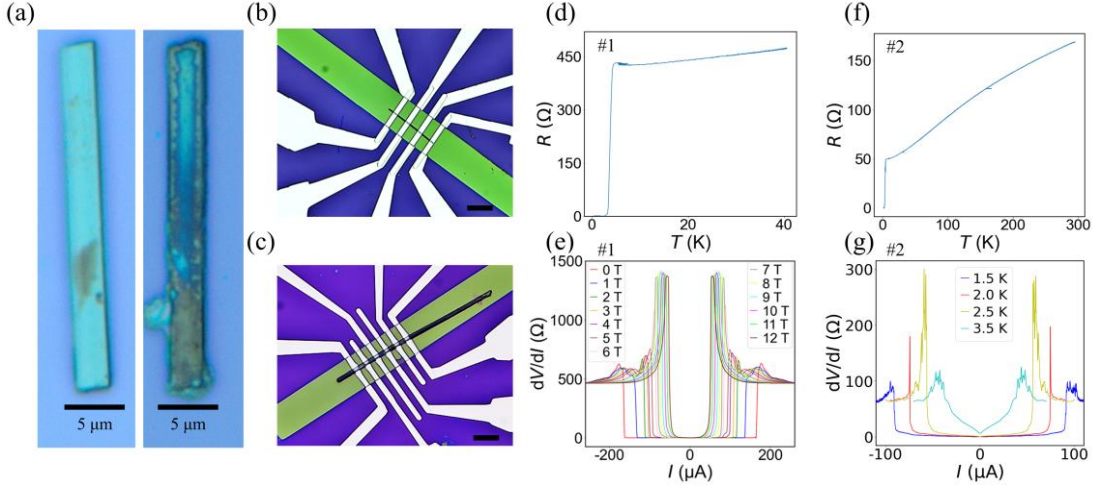

**FIG. 2.** $K_2Cr_3As_3$ devices. (a) Microscope images of $K_2Cr_3As_3$ before and after exposure to air. (b), (c) 1D application examples: $K_2Cr_3As_3$ devices with four and six electrodes, respectively. Scale bars: 10 μm. The white strips are Al electrodes, the green strip is the $Al_2O_3$ encapsulation layer, and the black wire in the middle is the $K_2Cr_3As_3$ sample. (d) $R$ versus $T$ for device #1. (e) $dV/dI$ versus $I$ under different magnetic fields for device #1. (f) $R$ versus $T$ for device #2. (g) $dV/dI$ versus $I$ under different temperatures for device #2.

$K_2Cr_3As_3$ has attracted considerable attention due to its unconventional superconducting properties, particularly its potential for spin-triplet pairing. Several key experimental findings have been observed, for instance, the absence of Hebel-Slichter coherence peak blow $T_c$ [14], the upper critical field exceeding the Pauli limit [24-26], linear temperature dependence of London penetration depth [19, 20], a suppression of spectral intensity approaching the Fermi level [27], the emergence of spin nematicity (broken rotational symmetry) [28], and possible $p$-wave pairing [23]. The unique structure of $K_2Cr_3As_3$ consists of 1D $(Cr_3As_3)^{2-}$ double-walled sub-nanotubes, which are separated by columns of $K^+$ ions. However, because of its reactive nature, $K_2Cr_3As_3$ is highly sensitive to environmental factors and is prone to rapid degradation. For example, even a brief exposure to ethanol can alter its structure [29], making it nearly impossible to process using conventional fabrication methods. Fig. 2(a) shows optical images of an exfoliated $K_2Cr_3As_3$ sample before and after a brief exposure to air, revealing significant surface degradation. In contrast, devices fabricated using our dry-transfer method, including those with four and six electrodes [Fig. 2(b), 2(c)], remain intact and free from damage. The material is encapsulated by an $Al_2O_3$ layer [Fig. 2(b), 2(c)], which is shown in green, and the Al electrodes are in silver color. Electrical measurements of these devices reveal a linear temperature dependence of resistivity [Fig. 2(d) for device #1 and Fig. 2(f) for device #2], consistent with bulk $K_2Cr_3As_3$ properties [20]. Furthermore, as shown in Fig. 2(e) and 2(g), the influence of magnetic field and temperature on the superconductivity of $K_2Cr_3As_3$ was examined. Our devices retain superconducting under a vertical magnetic field of 12 T, exceeding the Pauli limit, in agreement with previous reports on bulk materials [25, 26]. These results demonstrate that our method causes no damage on the 1D extremely sensitive material down to sub-micron scale and fully preserves its intrinsic properties.

$WTe_2$ is a layered transition-metal dichalcogenide that has garnered considerable attention due to its potential as a type-II Weyl semimetal [30, 31] and a candidate for higher-order topological insulators [32]. It's exceptional electronic properties, such as a large positive magnetoresistance which can be affected by drastic pressure, have made it a subject of intense study [33, 34]. The heterostructure of $WTe_2$ with other materials also leads to several novel phenomena, such as an ultrafast non-excitonic valley Hall effect [35], topological Hall effect in WTe$_2$/Fe$_3$GeTe$_2$ [36], and field-free switching of perpendicular magnetization in PtTe$_2$/WTe$_2$ [37]. However, WTe$_2$ is a sensitive material, e.g., O$_2$ can be chemically adsorbed and form defects [38].

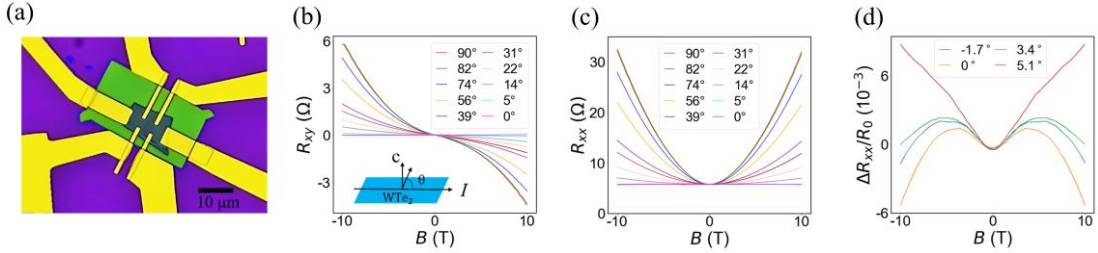

**FIG. 3.** WTe$_2$ device. (a) 2D application example: WTe$_2$ Hall-bar device. The yellow strips are Ti/Au electrodes, the green strip is the Al$_2$O$_3$ encapsulation layer, and the dark flake is WTe$_2$. (b) Angle-dependent transverse magnetoresistance $R_{xy}$ of WTe$_2$. (c) Non-saturating longitudinal magnetoresistance $R_{xx}$ in WTe$_2$. (d) Negative longitudinal magnetoresistance $R_{xx}$, which is apparently suppressed at ~5°.

To demonstrate the capabilities of our method, we fabricated a WTe$_2$ Hall-bar device, as shown in Fig. 3(a). The device is cooled to 0.3 K, and magnetic fields are applied in different directions to explore the magnetoresistance. When the magnetic field is aligned parallel to the c-axis ($B \perp I$, $B//c$), $\theta=90°$, while when $B$ is perpendicular to the c-axis ($B \perp c$, $B//I$), $\theta=90°$, as shown in the inset of Fig. 3(b). The angle dependence of the transverse magnetoresistance, $R_{xy}$, exhibits a characteristic trend that allows us to calibrate the angle $\theta$ [Fig. 3(b)]. In Fig. 3(c), we present the non-saturating positive longitudinal magnetoresistance, $R_{xx}$. When $B$ is aligned parallel to the c-axis ($\theta = 90°$), the positive magnetoresistance is maximized, and it diminishes significantly as $\theta$ decreases, eventually vanishing when $\theta$ reaches zero. This trend is fully consistent with the previous studies on $WTe_2$ [29]. In addition, we observed negative longitudinal magnetoresistance around $\theta = 0°$, as shown in Fig. 3(d), which is a signature of the topological properties. These results confirm that the intrinsic characteristics of $WTe_2$ are preserved during the fabrication process, demonstrating the effectiveness and broad applicability of our dry-transfer method for non-invasive fabrication of mesoscopic devices on sensitive 2D materials.

In conclusion, our dry-transfer fabrication method is especially suited for sensitive materials, as it completely avoids exposure to water, oxygen, and the solvents commonly used in traditional micro-fabrication processes. This approach is particularly beneficial for materials that are heat-sensitive. By utilizing this method, we can fabricate devices on both 1D and 2D materials, achieving high-quality electrical contacts, without strict restrictions on material thickness. The fabrication and characterization of $K_2Cr_3As_3$ and $WTe_2$ devices demonstrate the robustness and versatility of this technique. Our results show that this approach provides a reliable and efficient platform for the fabrication of mesoscopic devices on materials with exotic electronic, magnetic, and topological properties, opening up possibilities for advanced quantum devices and materials science research.

**SUPPLEMENTARY MATERIAL**

See supplementary material for additional figures and video.

**ACKNOWLEDGMENT**

We would like to thank Kun Jiang for fruitful discussions. This work was supported by the National Key Research and Development Program of China (2022YFA1403400); by the National Natural Science Foundation of China (12074417, 92065203, 92365207, U22A6005 and U2032204); by the Strategic Priority Research Program of Chinese Academy of Sciences (XDB33000000); by the Synergetic Extreme Condition User Facility sponsored by the National Development and Reform Commission; and by the Innovation Program for Quantum Science and Technology (2021ZD0302600).

**AUTHOR DECLARATIONS**

**Conflict of Interest**

The authors have no conflicts to disclose.

**Author Contributions**

**Zhongmou Jia:** Conceptualization (equal); Data curation (lead); Formal analysis (lead); Investigation (equal); Writing-original draft (lead). **Yiwen Ma:** Conceptualization (equal); Data curation (equal); Investigation (equal). **Zhongchen Xu:** Investigation (equal). **Xue Yang:** Data curation (equal); Formal analysis (equal). **Jianfei Xiao:** Data curation (equal); Software (equal). **Jiezhong He:** Data curation (equal); **Yunteng Shi:** Data curation (equal); Software (equal). **Zhiyuan Zhang:** Data curation (equal); Investigation (equal). **Duolin Wang:** Data curation (equal); Software (equal). **Sicheng Zhou:** Data curation (equal); Investigation (supporting). **Bingbing Tong:** Data curation (supporting). **Peiling Li:** Data curation (supporting). **Ziwei Dou:** Data curation (supporting); **Xiaohui Song:** Data curation (supporting). **Guangtong Liu:** Data curation (supporting). **Jie Shen**: Data curation (supporting). **Zhaozheng Lyu:** Data curation (supporting). **Youguo Shi:** Funding acquisition (equal); Investigation (equal). **Jiangping Hu:** Conceptualization (equal); Investigation (equal) ; Supervision (equal). **Li Lu**: Investigation (supporting); Funding acquisition (equal); Supervision (equal).

**Fanming Qu**: Conceptualization (equal); Funding acquisition (equal); Investigation (equal); Supervision (equal); Writing – review & editing (lead).

## DATA AVAILABILITY

The data that supports the findings of this study are available from the corresponding authors upon reasonable request.

# Supplementary Materials

# A non-invasive dry-transfer method for fabricating mesoscopic devices on sensitive materials

*Zhongmou Jia,* [1,2] *Yiwen Ma,* [1,2] *Zhongchen Xu,* [1,2] *Xue Yang,* [1,2] *Jianfei Xiao,* [1,2] *Jiezhong He,* [1,2] *Yunteng Shi,* [1,2] *Zhiyuan Zhang,* [1,2] *Duolin Wang,* [1,2] *Sicheng Zhou,* [1,2] *Bingbing Tong,* [1,3] *Peiling Li,* [1,3] *Ziwei Dou,* [1] *Xiaohui Song,* [1,3] *Guangtong Liu,* [1,3] *Jie Shen,* [1] *Zhaozheng Lyu,* [1,3] *Youguo Shi,* [1,2,4,a)] *Jiangping Hu,* [1,2, 5,a)] *Li Lu,* [1, 2,3,a)]

*and Fanming Qu* [1,2,3,a)]

**AFFILIATIONS**

[1] Beijing National Laboratory for Condensed Matter Physics, Institute of Physics, Chinese Academy of Sciences, Beijing 100190, China

[2] University of Chinese Academy of Sciences, Beijing 100049, China

[3] Hefei National Laboratory, Hefei 230088, China

[4] Songshan Lake Materials Laboratory, Dongguan, Guangdong 523808, China

[5] New Cornerstone Science Laboratory, Shenzhen 518054, China

[a)] Authors to whom correspondence should be addressed: ygshi@iphy.ac.cn; jphu@iphy.ac.cn; lilu@iphy.ac.cn; fanmingqu@iphy.ac.cn

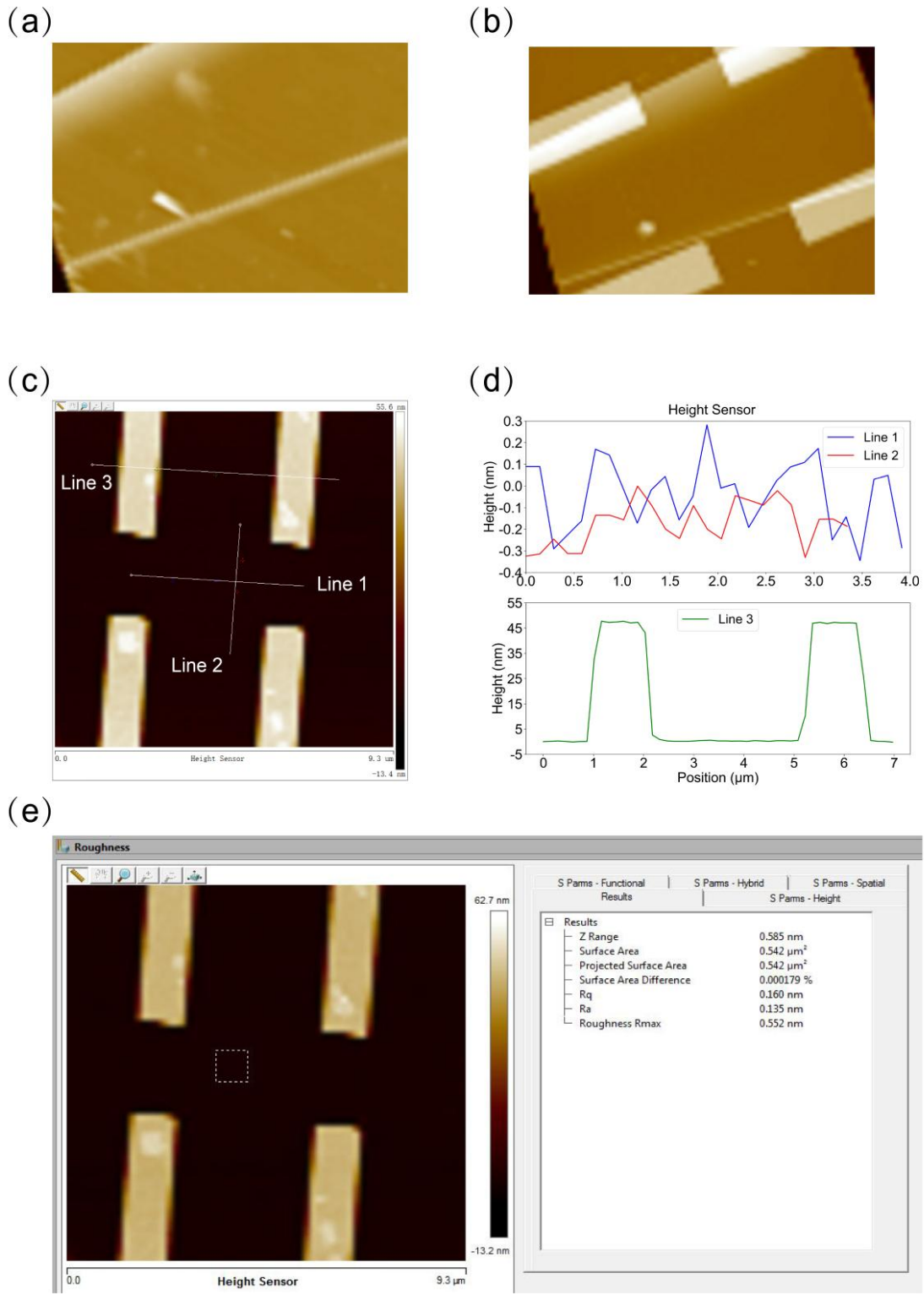

**FIG. S1.** AFM characterization of device. (a) Topography of pristine h-BN. (b) AFM morphology of the h-BN fabricated using our method, showing no new impurities introduced compared to pristine h-BN. (c) AFM morphology of the SiSiO$_2$ substrate fabricated using our method. (d) Height profiles of three lines in (c). The upper

subfigure shows height profiles along line 1 (blue) and line 2 (red), and the lower panel highlights height profile along line 3 (green). (e) RMS roughness of mask-substrate area ($R_q < 0.2$ nm).

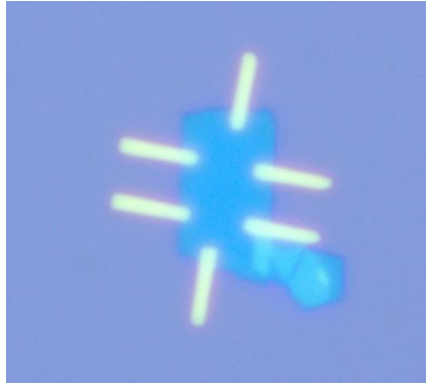

**FIG. S2.** Demonstration of the device fabricated with 10-nm-thick $WTe_2$. For thinner samples, we ecommend using a transfer stage system with higher-viscosity PDMS-assistdance to peel off metal film and mask.